\documentclass[sigconf]{acmart}

\renewcommand\footnotetextcopyrightpermission[1]{} 
\usepackage{booktabs} 
\usepackage{etoolbox}
\apptocmd{\thebibliography}{\raggedright}{}{}
\usepackage[utf8]{inputenc}
\usepackage[official]{eurosym}
\usepackage{array}
\usepackage{lipsum}
\usepackage{url}%
\usepackage[inline]{enumitem}

\begin{document}

\title[Understanding the challenges faced in complying with the GDPR]{Are we there yet? Understanding the challenges faced in complying with the General Data Protection Regulation (GDPR)}

\author{Sean Sirur} 
\affiliation{\institution{Department of Computer Science\\University of Oxford, UK}}
\email{sean.sirur@stx.ox.ac.uk}

\author{Jason R.C. Nurse} 
\affiliation{\institution{School of Computing\\University of Kent, UK}}
\email{j.r.c.nurse@kent.ac.uk}

\author{Helena Webb}
\affiliation{\institution{Department of Computer Science\\University of Oxford, UK}}
\email{helena.webb@cs.ox.ac.uk}

\renewcommand{\shortauthors}{Sirur, Nurse \& Webb}

\begin{abstract}

The EU General Data Protection Regulation (GDPR), enforced from 25\textsuperscript{th} May 2018, aims to reform how organisations view and control the personal data of private EU citizens. The scope of GDPR is somewhat unprecedented: it regulates every aspect of personal data handling, includes hefty potential penalties for non-compliance, and can prosecute  any company in the world that processes EU citizens' data. In this paper, we look behind the scenes to investigate the real challenges faced by organisations in engaging with the GDPR. This considers issues in working with the regulation, the implementation process, and how compliance is verified. Our research approach relies on literature but, more importantly, draws on detailed interviews with several organisations. Key findings include the fact that large organisations generally found GDPR compliance to be reasonable and doable. The same was found for small-to-medium organisations (SMEs/SMBs) that were highly security-oriented. SMEs with less focus on data protection struggled to make what they felt was a satisfactory attempt at compliance. The main issues faced in their compliance attempts emerged from: the sheer breadth of the regulation; questions around how to enact the qualitative recommendations of the regulation; and the need to map out the entirety of their complex data networks.

\end{abstract}

%
%
%

\copyrightyear{2018}
\acmYear{2018}
\setcopyright{acmcopyright}
\acmConference[MPS '18]{2nd International Workshop on Multimedia Privacy and Security at the 25th ACM Conference on Computer and Communications Security (CCS)}{October 15, 2018}{Toronto, ON, Canada}
\acmBooktitle{2nd International Workshop on Multimedia Privacy and Security (MPS '18) at the 25th ACM Conference on Computer and Communications Security (CCS), October 15, 2018, Toronto, ON, Canada}
\acmPrice{XXXX}
\acmDOI{XXXXXX}
\acmISBN{XXXXXX}

\ccsdesc[500]{Social and professional topics~Governmental regulations}
\ccsdesc[500]{Security and privacy~Privacy protections}
\ccsdesc[500]{Human- centered computing~Empirical studies in HCI}

\keywords{Data protection, regulations, GDPR, privacy, compliance, cyber security, multimedia, business, SMEs/SMBs}

\maketitle

\section{Introduction}
In recent years, data protection has become a forefront issue in cyber security. The issues introduced by recurring organisational data breaches, social media and the Internet of Things (IoT) have raised the stakes even further \cite{nursecurity,IOTDP1}. The EU General Data Protection Regulation (GDPR) \cite{GDPR}, enforced from 25\textsuperscript{th} May 2018,  is an attempt to address such data protection issues across all forms of media the world over. As a regulation designed to protect the privacy of EU citizens (regardless of the location of their data), the expectations of GDPR attempt to be as independent as possible of the complexity of the systems that it concerns. Every digital media format could be scrutinised under GDPR, from traditional software systems to the most cutting edge of distributed IoT technology.

The GDPR aims to incite a shift not only in the handling of personal data but in the attitude towards it as well. It gives EU courts power to severely punish any enterprise in the world that mistreats its citizens' data as according to the regulations. However, GDPR's concerns are not only punitive in nature. They also aim to provide a comprehensive template of the ideals of personal data protection (from an EU perspective) \cite{GDPR}. Chronologically, GDPR's life-cycle began with a proposal in 2012 which continued the trend in previous EU regulations of incorporating qualitative organisational issues rather than just prescribing technical controls \cite{Tankard}.

While undoubtedly advantageous in many ways for citizens and organisations, a reality with the GDPR is that organisations are having severe difficulties in understanding what compliance means in this new environment and how to implement it \cite{Agarwal,Tankard}. These data protection reforms are, for some, far beyond the scope of any previous instances \cite{COMPUK,T-Piri,ICOGUIDE}. The EU Council may be satisfied enough with GDPR to enforce it but the feasibility of complying with the regulations must be investigated. Developing and releasing successful data protection legislation is a continuous process where verifying the efficacy of the existing legislation helps to ensure that future legislation takes the best approach and feasibility is a necessity to any such success. This paper contributes to this process by enquiring into the issues faced by organisations when attempting compliance in the period around GDPR's implementation. 

The first objective is to understand the experiences of the organisations when interacting with the regulations: 
\begin{enumerate*}[label=(\roman*)]
\item 
investigating how feasible it was to translate the regulations into a technical context; 
\item
understanding the enterprise's perceptions of the regulations in terms of their ease of understanding; 
\item
gauging the organisations' awareness of GDPR and how this affected their compliance process; 
\item 
and the expectations the organisations had for the enforcement of GDPR. \end{enumerate*} 
The second objective is to study the processes used by organisations when implementing GDPR: 
\begin{enumerate*}[label=(\roman*)]
\item
what mechanisms and techniques were employed during implementation; 
\item
what support they sought from governmental, industrial and academic sources; 
\item
and how they went about verifying their compliance.
\end{enumerate*} 

This study found that, while organisations felt compliance was doable, the compliance attempts of large companies and SMEs focused on data protection-focused were of a higher maturity than that of more general SMEs. Also, compliance took at times a drastic toll on the latters' resources. This resulted in a rift in the attitudes of the two groups towards compliance expectations and feasibility. The structure of the paper is as follows: Section~\ref{pw} describes existing work; Section~\ref{me} presents the paper's methodology; Section~\ref{re} describes the results of the study; Section~\ref{cl} reflects upon the study itself and suggests further work. 

\section{Previous Work}
\label{pw}

GDPR by its very nature has attracted much attention from industry and academics alike. Articles and opinion pieces discussing it are fairly numerous. Due to its relative modernity, however, there is no vast body of academic literature present. In what follows, we aim to reflect on the most relevant existing research. 

\subsection{GDPR Awareness}

A key interest in both the academic and industrial communities is the matter of awareness. Two studies attempted to gauge the reality of organisations' awareness empirically. An early 2014 academic study performed in Finland by Mikkonen spoke to likely data controllers. Data processors and data controllers as defined in GDPR are both enterprises that process data but data controllers have the responsibility of ensuring that data is protected as per GDPR \cite{ICOGUIDE4}). The study showed that 43\% of data controllers were aware of GDPR with only around three-quarters of that group (i.e. 31\% of all data controllers) willing to act on GDPR at the time of the study \cite{Mikkonen}. This study was performed closer to the beginning of the GDPR lifespan hence the uncertainty present in the actions of the companies.

Despite the imminence of GDPR in the approach to 2018, a joint industry and academic 2017 UK study of business and charity awareness (Cyber Security Breaches 2018) showed alarmingly similar numbers after three years: 38\% business awareness and 44\% charity awareness. The numbers drop further when depth of knowledge is questioned \cite{UKSTUDY}. It must be noted, however, that the results of the two studies are not necessarily directly comparable. The UK study does not specify that they solely spoke to data controllers but rather an assortment of businesses and charities as a whole. 

The UK study also had interesting findings on the most common changes made by organisations surveyed in the lead up to GDPR. By quite a large margin, the most changes for both demographics was in policies and procedures. This is understandable as, given GDPR's focus on accountability, nearly every change would result in some policy or procedure modification. For businesses, additional staff training and communications almost doubled the activity levels of any other change, except policy and procedure, comprising a fifth of the most common changes. 

\subsection{Preparing for the GDPR}

The organisations' reactions as described in the UK study were largely predicted in a Finnish paper discussing the implications of GDPR \cite{T-Piri}. It outlines many likely steps that organisations will have to take to ensure a robust level of GDPR compliance. The paper methodically constructs a selection of implications concerning various matters such as consent acquisition, documentation maintenance and data protection by design. It thoroughly discusses each implication alongside potential areas of difficulty. It avoids giving direct advice on how to overcome issues, however. Regardless, the material present is a thorough and clear deconstruction of GDPR's expectations.

Academic research also focuses on the issues around privacy in the Internet of Things (IoT). One early Finnish legal research article \cite{IOTDP1} described the IoT issues stemming from the wide deployment of IoT devices outside of traditional technology environments as well as the dangers of the lack of built-in security in many IoT devices. The article stakes its hopes on GDPR curbing what was at the time an ever increasing consumption of data by IoT vendors and other related enterprises. The author focuses on the proposed regulation's focus on transparency and data minimisation principles, stating that accountability could be a key feature of future data protection laws. The article also considers that a risk-based approach when differentiating sensitive and non-sensitive data may be necessary because, according to the author, IoT devices often process sensitive data illegally but unknowingly. 

Such risks are reinforced by users' ignorance and apathy towards privacy concerns: ``the privacy paradox'' \cite{willperfect,OX1}. Attacks through IoT devices have been demonstrated that greatly challenge the average users' view of privacy \cite{OX2}. This serves to show how privacy concerns cannot be relegated solely to the user base of a product and that the developers and vendors of such goods and services must be held accountable for. 

The discussion within the above Finnish legal research article is reassessed by a more recent 2018 article \cite{IOTDP2}, also from Finland, in light of GDPR's announcement. The author is confident that GDPR will regulate IoT practices in a way that the previous EU Data Protection Directive (DPD) failed to do and that it generally appears to be a good move towards improving user privacy. However, the article outlines two future risks: firstly, the perceived difficulty for SMEs to implement the lengthy and numerous articles in GDPR and, secondly, that the qualitative aspects of GDPR allow too much flexibility, failing to push the status quo far enough in the right direction. 

The 2\textsuperscript{nd} GDPR Readiness Survey was an in-depth study into organisational compliance in the months before GDPR's enforcement \cite{CIPL}. The study found that budgets for GDPR compliance could approach \$50 million for a single company. Also, despite an emphasis on improving technical systems, data mapping and tagging was primarily a manual process. Thirdly, the exact meaning of certain terms in GDPR were still in need of clarification. An additional academic analysis of the results found that the three major concerns appeared to be privacy management, handling third-party data processors and the data breach disclosure. The majority of respondents were multi-national corporations.

An American law firm conducted research into GDPR preparations for FTSE 350 and Fortune 500 firms. Their survey concluded that major companies have assigned large budgets to GDPR compliance (with large portions allocated to technology) but that this still may not be enough \cite{PH}. The Fortune corporations (United States) spent a little under double of the FTSE companies (London Stock Exchange) on average. The research claims that individual budgets set aside for additional permanent staff alone were upwards of £200,000 for 40\% of the FTSE 350 and greater than \$500,000 for the Fortune 100 despite only 10\% of both UK and US firms having bought new technology. It must be noted that the study did not include any SMEs.

The most recent report on a study conducted across the EU, UK and US \cite{COMPLIANCE} found that only a fifth of organisations considered themselves fully compliant. Around 27\% of companies had not begun the compliance process with remaining companies being in the midst of complying. The primary difficulty cited was the complexity of GDPR. The budgets expended on GDPR compliance were large, ranging from \$500,000 to over a million dollars. 

\subsection{Research Motivation}

Though many organisations and individuals understand the gravity of the regulations, uncertainty around GDPR's nuances has led to a somewhat divided approach, destabilising GDPR as a unifying set of data protection rules. GDPR aims to centralise and demystify data protection practices but it is mired in debate, though that may be a necessary part of its evolution. 

While there is an established body of research on GDPR and legislative data protection as a whole, the regulation's novelty means that no deep qualitative studies into organisational compliance exist. Most studies of real companies appear to focus on reporting either compliance and awareness percentages or, for example, concrete metrics such as the budget sizes expended on compliance or the increase in workforce size. There is yet to be in-depth qualitative research focusing on the perceptions and experiences of organisations in their attempts to comply with GDPR. This study addresses that gap.

\section{Methodology}
\label{me}

The aim of our research study is to gain an in-depth understanding of the real challenges and concerns faced by organisations in complying with the GDPR. In particular, the study focused on answering the following research questions:

\begin{enumerate}

\item

What did those who worked directly with GDPR think of the document itself?

\item

What was the process of implementing GDPR compliance like for these organisations?

\end{enumerate}

Interviews were chosen as the appropriate method to address the research questions. A set of interview questions was defined that covered all the areas of interest while being flexible enough to cope with the various backgrounds of the respondents. The interview questions were generated from the primary research questions, with previous research being used as inspiration for question topics \cite{UKSTUDY} \cite{T-Piri}. This also gave an opportunity to validate previous findings. The UK ICO GDPR guidelines, particularly the self-assessment tool-kit, helped gauge what questions the ICO expected organisations to answer \cite{ICOGUIDE3}.

The interviews were semi-structured to allow the questions to flow naturally with the expertise of the interviewees whilst still acquiring answers from consistent themes. As a result, each of the three main topics were introduced with an intentionally non-specific question, allowing the interviewees to speak freely about the area. This gave the interviewer the opportunity to gauge the level of interest and expertise each of the interviewees had for the various topics. The questions aimed to be as neutral as possible. New topics were introduced broadly e.g. \textit{``tell me about the processes you used to implement compliance"}. This prevented the interviewees' focus from being led away from their own priorities. More focused questions avoided emphasising a bias for or against any response. Questions that were predicted to be more esoteric or non-specific had examples available. The process followed guidance in \cite{Qual,Qual2}.

Best ethics practice was followed and clearance acquired from the authors' ethical research regulatory body. Candidates were recruited by various means including utilising professional contacts in the authors' practitioner and research communities or through snowball sampling. Interviews were conducted either in person, over the phone or via internet voice chat. Consent was explicitly requested and documented from candidates prior to the interview and recording process commencing.

The interviews were analysed in two stages: \textit{familiarisation} and \textit{thematic coding} \cite{Qual3}. In the first stage, anonymised transcripts were manually familiarised: recurring themes were written down and sorted by their topic and relation to the research questions. In the second stage, the transcriptions underwent qualitative coding analysis. The primary research objectives and the themes found in the previous step were combined into a hybrid coding frame. Then, sections of each transcript were tagged with relevant codes to aid a more thorough understanding.

\section{Results and analysis}
\label{re}
The study attracted respondents from a variety of different backgrounds working in different areas. This was useful as it allowed a more balanced perspective of the attempts at GDPR compliance and the issues faced. Respondents are outlined in Table~\ref{tab1}.

\begin{footnotesize}
\vspace{2mm}
\begin{table}[h]
\renewcommand{\arraystretch}{1.1}
\begin{center}
\begin{tabular} { | m{1.2cm} m{4cm} m{2cm} | }
  \hline
  \textbf{Respon-dent} & \textbf{Role} & \textbf{Years of cyber security experience} \\
  \hline
  \hline
  R1 & University Senior Security Executive (Technical) & 24 \\
  \hline
  R2 & Fortune 500 Senior Security Executive (Technical) & 20 \\
  \hline
  R3 & Privacy and Data Protection Consultancy (SME) CEO & 10 \\
  \hline
  R4 & Previous Security Hardware Solutions Company CEO (SME), Security Researcher and current Entrepreneur & 15-20 \\
  \hline
  R5 & Cyber-threat Intelligence Company (SME) Senior Security Executive (Governance) & 3 (8 years in law enforcement security analytics) \\
  \hline
  R6 & Information Security and Penetration Testing (SME) Consultancy Senior Security Executive (Governance) & 14 \\
  \hline
  R7 & Fortune 500 Technology Company Senior Security Executive (Technical) & 28 \\
  \hline
  R8 & Government Sector Senior Security Executive (Governance) & 8 \\
  \hline
  R9 & Telecommunications Senior Executive and Freelance Consultant (Governance and Development) & 25 \\
  \hline
  R10 & Data Software and Research Company (SME) Senior Executive (Law and Policy) & 20	\\
  \hline
  R11 & Technology and Software Development Company (SME) Senior Executive (Human Resources) & 20+ \\
  \hline
  R12 & Internet Development Non-Profit (large organisation) Senior Executive (Policy and Strategy) & 15 \\
  \hline
  
\end{tabular}
\vspace{2mm}
\caption{Details of Study Participants}
\label{tab1}
\end{center}
\end{table}
\vspace{0mm}
\end{footnotesize}

All but one respondent (R4) had considerable experience working directly with implementing GDPR. Those who had worked directly with GDPR felt that, whilst arduous, compliance was a worthy endeavour. Each had varying levels of comfort regarding their organisations compliance level. The results are structured by the main themes found in the study.

\subsection{Experiences with the Regulations}

\textbf{\\ Translating GDPR into a Technical Context} \\

\setlength{\leftskip}{0.4cm} \setlength{\rightskip}{0.4cm}

\noindent \textit{``On the face of it is very easy to read and understand but... there's a lot of it. You have to think about what does this actually mean to [my organisation]" \\} --- Respondent R1 \\

\noindent \textit{`` It's probably too much work for the average engineer... the debate starts immediately if something technical is rising up..." \\} --- Respondent R7 \\


\setlength{\leftskip}{0cm} \setlength{\rightskip}{0cm}

The challenges in translating GDPR into a technically implementable format came from a range of issues. While GDPR was feasible to implement for large organisations with the necessary resources,  smaller organisations were not always in this position. The largest non-technical issue for most respondents was in deciphering the expectations of GDPR itself (or the corresponding state-issued guidance). The respondents from larger organisations and security-focused SMEs/SMBs expressed a sense of satisfaction with the process as a whole, claiming that the effort had given them a firmer understanding of their company's data protection stance. For the other SMEs/SMBs, considerable effort had to be expended to understand what was expected of their organisations to comply. They were at times unable to comment on the benefit of GDPR's broader philosophies due to the major hurdles faced in basic compliance.

Some respondents discussed GDPR's generality, which did not lend itself to  concrete technical requirements. These qualitative aspects of GDPR appeared to cause some of the most issues. However, all interviewees agreed that GDPR was a step towards more thoughtful cyber security practices. Multiple respondents felt that these areas of GDPR expected some form of risk management. This was used as an opportunity to reassess their organisation's stance towards cyber-risks and threats with data protection impact assessments holding a vital role in this process. While the more confident organisations praised GDPR's use of qualitative statements to account for the complexities of data protection, such passages were a major stress on SMEs when beginning the process of GDPR compliance. For larger organisations, the qualitative areas signified a tone of flexibility and nuance but for some SMEs it appeared as a lack of clarity.

\textbf{\\ Readability and ease of understanding} \\

\setlength{\leftskip}{0.4cm} \setlength{\rightskip}{0.4cm}

\noindent \textit{``On the face of it is very easy to read and understand but... there's a lot of it." \\} --- Respondent R1 \\

\noindent \textit{``I think `appropriate' helps cover a lot of ground. I don't think it is a bad phrase, it requires you to think about what your risk is." \\} --- Respondent R2 \\
	
\setlength{\leftskip}{0cm} \setlength{\rightskip}{0cm}

The respondents who directly worked with the GDPR document itself felt it was written with clarity, stating that it was evident care had gone into its construction. One promising recurring theme was that most of the respondents felt that GDPR was a readable and digestible document despite many of them being technical or managerial experts, not legal specialists.

Understanding the semantics and meaning behind the words of GDPR was not as simple. Despite their thoughts on the clarity of GDPR's intentions, the more technically focused respondents also expressed the sentiment that without a legal professional of some kind the average engineer would struggle to utilise the regulations directly. Respondents with a mixed background stated that analysing and understanding GDPR was doable but challenging. The interviewees with a stronger legal and policy understanding were divided: GDPR was, idealistically, a definite improvement but was far from the standard necessary to truly ensure holistically safe data practices.

\textbf{\\ Awareness} \\

\setlength{\leftskip}{0.4cm} \setlength{\rightskip}{0.4cm}

\noindent \textit{``Now that a lot of companies are thinking about how they're processing data and how they're storing it... that's a good thing!" \\} --- Respondent R10 \\

\noindent \textit{``The size of GDPR looks very daunting and when people with limited capacity of any kind see this they can default to not doing it." \\} --- Respondent R9 \\

\setlength{\leftskip}{0cm} \setlength{\rightskip}{0cm}

The responses varied quite widely on this matter, highlighting the disparity in the level of awareness between organisations that were security-focused and those that were not. All of the companies involved in this study were aware of GDPR and made steps to begin their compliance processes before the deadline. Of those organisations that were not security-focused, however, their respondents stated that the compliance process was, at least initially, driven largely by their own endeavours.

The reason for failures to consider GDPR are not well explained. It appears that while security-conscious organisations have been aware of GDPR for a relatively long time, other organisations and individuals have not successfully recognised GDPR as a priority. Ideally, companies have had a few years to prepare for GDPR but this preparation time may have gone to waste if too many companies were either unaware of GDPR or had just started working on compliance around the deadline. The alleged delay in the ICO's guidance (as mentioned by multiple respondents) may be related to this issue. In addition to this, some SMEs/SMBs that were aware of GDPR were unable to make substantive steps towards specific GDPR compliance, despite anxiety around the issue, due to strained resources and a lack of guidance from the ICO until relatively close to the deadline.

\textbf{\\ Expectations of Judicial Response} \\

\setlength{\leftskip}{0.4cm} \setlength{\rightskip}{0.4cm}

\noindent \textit{``What legislation trumps GDPR or are trumped by GDPR? [...] In some cases, they totally exclude each other."  \\} --- Respondent R8 \\

\noindent \textit{``Most of these things are going to be a negotiation around the table. Your lawyers are going to [debate] and agree on something." \\} --- Respondent R4  \\

\noindent \textit{``A lot of it will depend on the interpretation prior to and after the first case of GDPR. Appropriateness is in the eye of the beholder."  \\} --- Respondent R2 \\

\setlength{\leftskip}{0cm} \setlength{\rightskip}{0cm}

On this topic the entire set of respondents was in agreement: for GDPR to have a constructive impact, the courts would have to be pragmatic and reasonable in their enforcement of GDPR. All the organisations who participated in the interview made it clear that their attempts at compliance were under the assumption that GDPR would be used constructively. The general consensus here was that, despite anyone's best efforts at compliance, the reality will only be apparent after the prosecution of some cases. The concern for most organisations (even the larger ones) was that, despite their best efforts, the ICO may have a different perspective on responsible practice and as a result may penalise them.

However, given the Information Commissioner's statement on their desire to sensibly enforce GDPR \cite{ICO2}, pragmatism seems to be a priority. This was reflected in the smaller UK organisations' heavy reliance on the guidance of the ICO (which is itself the UK's data protection regulator). Unfortunately, the latter point was still a source of friction for many respondents as, despite their trust in the ICO, the belated nature of the guidance was a major cause for concern.

\subsection{The Implementation Process and Issues}

\textbf{\\ Mechanisms, Techniques and Processes} \\

\setlength{\leftskip}{0.4cm} \setlength{\rightskip}{0.4cm}

\noindent \textit{``There are mechanisms that we can use to apply technology to resolve GDPR" \\} --- Respondent R1 \\

\noindent \textit{``Risk management was a big focus" \\} --- Respondent R11 \\

\noindent \textit{``Data flow diagrams are kind of fundamental. Ultimately everything in IT is about data flowing from one place to another, being processed and stored." \\} --- Respondent R1 \\

\noindent \textit{``I love to put people in training. It is super important for anything" \\} --- Respondent R7 \\

\setlength{\leftskip}{0cm} \setlength{\rightskip}{0cm}

While GDPR introduced some new challenges, the technical abilities required to comply were not far from previous data protection standards. Respondents did not feel concerned about their existing development paradigms. Instead, the main topics discussed were data flow mapping, automated monitoring, protocol updates and training.

Data flow mapping was key in any compliance attempt. Knowing the behaviour of their data was imperative for an organisation to have robust data protection. Without understanding where their data was transmitted and stored, organisations felt they could not hope to have enough control over their data to protect it. While this was feasible for larger or more data protection-focused companies, this was a highly challenging task for most SMEs/SMBs due to the overhead involved in mapping out complex webs of data networks. Respondent R4, in particular, was sceptical of the accuracy present in any data networking map, using the failed attempts of a large technology firm with which the respondent was acquainted as an example.

When successfully implemented, the improved understanding from mapping data helped unearth risky data practices that were hitherto unseen. It was reported that talking to organisation employees and discussing the importance of respecting personal data was also effective to this end. The discourse apparently served in particular to highlight instances of employees who broke good data protection practice through benign ignorance as opposed to malice or carelessness. This highlights an issue for SMEs/SMBs which may not have the resources to carry out large-scale discussions/training and thus would remain at risk.

Another technical focus was the use of automation which could greatly facilitate technical compliance. The general use of automation seemed to be restricted to advanced monitoring, tagging and warning services, however. Automation was neither trusted nor effective enough to make anything more than rudimentary decisions around data protection as any error in the system could risk damaging the organisations' finances and reputation.

One ubiquitous non-technical response was regarding the increase in training and education. Many of the respondents stated that their organisation put many if not all employees through mandatory data protection awareness seminars. One respondent, R7, felt that good training trumped rigid and verbose engineering practices as well-trained engineers could make more flexible, informed and innovative decisions compared to a static software engineering model. Again however, SMEs/SMBs were unable to implement retraining at such a large scale, opting instead, for shorter seminars and workshops.

There appears to no simple way to address these issues. Much of the difficulty arises from the complex and unique nature of data systems: broadly speaking, no two systems are the same making every unique attempt at GDPR compliance necessarily non-trivial. Respondent R12 interestingly suggests that there is a possibility this will result in SMEs/SMBs opting to use (e.g. cloud) services provided by larger companies when possible as opposed to constructing or providing their own.

\textbf{\\ Support from government, industry and research} \\

\setlength{\leftskip}{0.4cm} \setlength{\rightskip}{0.4cm}

\noindent \textit{"I called them about 3 times for clarifications... [but] I think the ICOs guidance was very clear." \\} --- Respondent R5 \\

\noindent \textit{"It was the ICO's documentation and guidance which I found the most user friendly and relevant." \\} --- Respondent R11 \\

\setlength{\leftskip}{0cm} \setlength{\rightskip}{0cm}

Each organisation sought different levels of support from different environments. The most confident respondents generally did not rely on state support. For the largest companies, a single state's guidance may not be appropriate when operating in multiple legal jurisdictions. Furthermore, in the case of data protection forerunners such as R2, R3 and R7, industrial support would likely be generated by them as opposed to being sought. No organisation made use of academic research in their data compliance processes.

Smaller organisations benefited strongly from state support in the UK, making liberal use of the ICO's guidance and documentation. Though many of their GDPR compliance processes began some time before the ICO released any substantial guidance, this is in part implied to be more due to the ICO's lateness than due to their organisations' timeliness. It was agreed that, whilst governmental support was not entirely holistic in terms of applicability, the parts that were could be very effective. In addition, the ICO was reported as being timely when responding to queries and clarifications.

Respondent R4 was somewhat disparaging of the outcome of the support they expected from the government, stating that the ICO's proposed personnel increase was too small to handle the workload the entrepreneur expected GDPR to generate. However, they did state that the public attention brought to data protection by the ICO would be socially beneficial in the long term. Aside from this, the respondents were generally quite positive about governmental support. However, whilst the only strong critique was the lateness of the guidance, this was a major concern for smaller SMEs/SMBs that lacked a business focus on data protection as they were not equipped to tackle the GDPR regulation directly alone.  \\

\setlength{\leftskip}{0.4cm} \setlength{\rightskip}{0.4cm}

\noindent \textit{``We should be aware of the possibility that [SMEs] will prefer to channel some of their services through big companies which they're certain have the money to comply." \\} --- Respondent R12 \\

\noindent \textit{``[There are] people did a very small course of a couple of days, call themselves experts on GDPR and provide all kinds of incorrect advice... \\} --- Respondent R3 \\

\setlength{\leftskip}{0cm} \setlength{\rightskip}{0cm}

Several respondents remarked on the presence of predatory parties. Reports of non-reputable lawyers and technology companies offering their services regarding data protection was met with frustration by the entire set of interviewees. This hazard affects SMEs/SMBs the most. Combined with their present lack of resources, smaller organisations may fail to avoid fraudulent offers from predatory enterprises due to a lack of critical knowledge. There was no clear mechanism to prevent this and it is unknown how many organisations could have been affected this way, though none in the study were.

Positive feedback regarding support from the legal industry was lacking. No participant successfully cooperated with any legal professionals as part of their compliance process. The penetration testing company (R6) was sent a lawyer but apparently the lawyer appeared to have a weaker understanding of GDPR than those already present in the company. It is uncertain whether the presence of legal help would have been of aid. Respondent R7, however, felt legal experts had a place in the process. \\

\setlength{\leftskip}{0.4cm} \setlength{\rightskip}{0.4cm}

\noindent \textit{``I guess if someone sent me something it would be different but when I wanted to see if we were compliant in an area I want a yes or no answer so I'd look at the ICO website or calling them." \\} --- Respondent R5 on the direct applicability of academic research for SMEs/SMBs \\

\setlength{\leftskip}{0cm} \setlength{\rightskip}{0cm}

None of the companies utilised academic research in their compliance process with most finding academic research not pragmatic or transparent enough for immediate use. One respondent, R5, specified that there was not enough time to delve into research partially due its esoteric nature, particularly when the ICO's guidance itself was available. However, whether the applicability of the research or the method of dissemination was the primary issue in most cases was unclear.

\textbf{\\ Compliance and Verification} \\

\setlength{\leftskip}{0.4cm} \setlength{\rightskip}{0.4cm}

\noindent \textit{``I don't think it's just feasible, I'm certain that [our organisation] have done it."
\\} --- Respondent R2 \\

\noindent \textit{``When we started to discuss this 3 years ago the impression I got was that we have been compliant the whole time [...] people who care about privacy have been very close if not already compliant" \\} --- Respondent R7 \\

\noindent \textit{"The GDPR is a complex statute involving significant costs of compliance. It might be the case that it will be the big companies that find it easier to comply." \\} --- Respondent R12 \\

\noindent \textit{``We'd been looking after our data pretty well [...] we realised we're processing data in a lot of ways that never hit the radar before but we're dealing with it." \\} --- Respondent R5 \\

\noindent \textit{``Yes, we meet the minimum standards but I have better aspirations for the long run." \\} --- Respondent R8 \\

\noindent \textit{``Regardless of the size of the company [...] nobody's going to be fully compliant." \\} --- Respondent R10 \\

\noindent \textit{"There are things in the new GDPR legislation that I think would be very hard for companies to say they comply on." \\} --- Respondent R11 \\

\setlength{\leftskip}{0cm} \setlength{\rightskip}{0cm}

The quotes above illustrate a key finding of this study: the difference in language, tone and perceptions when discussing GDPR compliance between two organisation categories. The first (large organisations and security-focused SMEs), while pragmatic about the work involved in maintaining compliance, were generally confident about the outcome of their compliance process. The second (all the other SMEs) were usually satisfied with their attempts though not as often and with more attention towards the ``fuzziness'' around compliance.

Largely, there were several common themes around the respondents' views on compliance verification: accountability (records and audit trails); flexibility of the regulations; data mapping; and monitoring. All respondents directly responsible for GDPR compliance agreed that accountability via clear record trails went a long way to improve overall data protection standards. Whether the auditing and record keeping was of the same quality across the different organisations could not be inferred.

Another theme concerned the qualitative statements in GDPR: some felt qualitative statements showed that compliance expectations were not overly rigid but rather a push for organisations to undertake a more thoughtful data protection process. However, respondent R11 expressed fears that companies may use such statements to exploit the regulations but was unable to give more details. Regardless, as illustrated prior, qualitative statements were not comforting for every respondent. The recurring issues around translation and clarification faced by the less data protection-focused SMEs/SMBs could dwarf the benefits given by the flexibility.

Respondent R4 was, as mentioned above, highly sceptical of GDPR compliance and the notion that anyone had achieved it. They expressed the somewhat radical view that companies only complied for the sake of \textit{``fashion and explicitly enforced regulation"}. In their view, while some companies indeed tried to comply in earnest, many companies were just ticking boxes for auditors. This again ties in strongly with their experience as a businessperson that companies are largely profit driven. In R4's opinion, the average corporation would not attempt to achieve a data protection maturity level above that which minimises the risk of financial losses in the event of a data breach. This is, as a result, highly dependent on the judicial reaction to data protection failures and how harshly they are willing to penalise guilty parties.

Some of the views expressed by R4 are present in the results of this study, though perhaps not following quite the same narrative. As has been seen, smaller SMEs/SMBs genuinely did have to carefully measure what areas of GDPR compliance were most pressing for them given their extremely restrictive budgets and resources. Regardless of whether some companies wilfully shirk data protection responsibilities, it appears to be an unfortunate truth that some willing smaller organisations are still genuinely too ill-equipped to substantially comply with GDPR in the short term.

\section{Conclusion}
\label{cl}
Despite doubts around the introduction of GDPR, some organisations are satisfied with their compliance upon the deadline. Unfortunately, some of the fears around compliance have been realised with smaller companies struggling to keep up unless already focused on security. Those respondents belonging to organisations all found compliance doable, though not necessarily feasible in the allotted time-scale. In terms of the issues faced when complying, emphasis was placed on understanding the qualitative expectations of GDPR with respect to real systems; mapping organisational data flow; and verifying compliance without preceding case law for context. Most respondents stressed the benefits of state-issued guidance; training and education; record keeping; and (to some extent) automated monitoring. The results also showed that predictions made about the processes required to comply were accurate \cite{T-Piri}; the difficulties faced by SMEs in compliace \cite{IOTDP2}; and the budgets expended for GDPR compliance \cite{PH}.

While the interviewees supported responsible data protection, many aspects of GDPR created a disparity between the reactions of organisations with different levels of resources. Much of the confidence around GDPR compliance rests on the assumption that it will be assessed pragmatically. To many, it is up to the courts to draw the line between GDPR being pragmatic and beneficial or bureaucratic and stifling.

On the technical side, existing software engineering paradigms appear capable of handling GDPR. It must be noted that the feedback on this area was limited as respondents were not always versed in the technical aspects of compliance. Nevertheless, secure software development is a field which GDPR could still heavily influence. While software engineering processes were doable, feasibility in short time-scales was not guaranteed. It is possible that GDPR's introduction could result in large companies with high data protection capabilities acting as data-service providers (e.g. of cloud or database services) to SMEs.

Many felt that GDPR was a step in the right direction, finally giving companies an incentive to step back and inspect their data protection stance. The regulation appears to give a  freedom of interpretation to skilled individuals when considering compliance. However, this same freedom of interpretation risked alienating smaller companies which did not have the resources necessary to cope with the resulting overhead.

Unravelling and contextualising the regulations' qualitative statements was something that not all organisations could easily handle. Indeed, this is where external support was key, with governmental support performing particularly well in this role. Whilst government support programmes are already available, industry involvement could take the shape of training services, bespoke data protection services or even through the emerging role of cyber insurance ~\cite{baercyberinsurance,woodsmapping}. For research, understanding why and how some organisations succeeded in complying is imperative to improving and easing the introduction of new regulations. This understanding is also key to helping smaller organisations undertake data protection practices of a high maturity despite restrictions on resources. Only through addressing the issues faced by SMEs can robust data protection be made truly ubiquitous.

\section{Limitations and Further Work}

In this article, we presented an early study towards a holistic understanding of the non-academic reaction to GDPR in the wake of its enforcement. Care was taken to ensure the study was unbiased when constructing and conducting interviews and also throughout data analysis. Inevitably, each stage of the research process still carries the risk of introducing different issues into the results. Extending the study to include a longer time-frame and more participants would reduce many of the existing risks.

One possible source of skewing comes from the interview recruitment process. There was a risk that organisations willing to discuss experiences around GDPR were those with a more positive perception of their own compliance. Companies that felt uneasy with their level of compliance could very well be unwilling to discuss the matter with third-parties, particularly out of fear of the resulting repercussions or reputation damage. By its very nature, this skew would be very difficult if not impossible to avoid or even account for. Offering an incentive may go some way to alleviate this problem but it is unlikely to overcome any strong reluctance on the part of potential interview candidates. Another potential skew in the recruitment process comes from the majority of the interviewees being within the professional and academic circles around the authors. As a result, the average level of security and privacy maturity of the organisations contacted may have been non-representative of typical organisations. This is quite likely given the high level of compliance in this data pool versus the average level of compliance reported across the UK \cite{COMPLIANCE}.

The respondents' knowledge of different aspects of compliance varied, particularly around the deep technical aspects. For instance, some respondents struggled to understand questions focusing on privacy-oriented software engineering models and secure programming. This diminished the pool of technical responses. A further study focusing on talking solely to developers, programmers and engineers could help in understanding the issues faced when engineering compliant systems.

Another possibility is to conduct ensuing studies on a more focused interest, such as focusing entirely on the technical issues or the influence of industry support versus government guidance. Depending on the desired results and the precision of the study, a fully-structured interview process may be better for deriving quantitatively comparable results. This could be used, for example, to help measure the cost-effectiveness of budgeting different sectors (e.g. personnel training versus database upgrades) within an organisation to understand what changes contributed the most to improving compliance.
Studying the processes used by (arguably) compliant companies and trying to understand the key aspects that contributed to their compliance could go some way towards developing more effective models to streamline the design, development and maintenance of privacy-oriented systems for organisations who are less equipped to design such systems from the ground up.

\bibliographystyle{ACM-Reference-Format}
\bibliography{mps-submission-sirur-nurse-webb}

\end{document}